\documentclass[twocolumn,epjc3]{svjour3}

\RequirePackage{graphicx} 
\usepackage{bm}
\usepackage{epsfig}
\usepackage{subfig} 
\usepackage{multirow}
\usepackage[colorlinks=true,citecolor=blue,filecolor=blue,linkcolor=blue,urlcolor=blue,pdftex]{hyperref}
\usepackage{mathtools}
\usepackage{supertabular,booktabs}
\usepackage{lineno}
\usepackage[usenames,dvipsnames]{xcolor}
\usepackage[normalem]{ulem}

\definecolor{forestgreen}{rgb}{0.13, 0.55, 0.13}

\journalname{Eur. Phys. J. C}

\begin{document}
\title{A Dual-phase Xenon TPC for Scintillation and Ionisation Yield Measurements in Liquid Xenon}

\author{Laura~Baudis\thanksref{e1,addr1}
\and Yanina~Biondi\thanksref{addr1}
\and Chiara~Capelli\thanksref{addr1}
\and Michelle~Galloway\thanksref{addr1}
\and Shingo~Kazama\thanksref{addr1}
\and Alexander~Kish\thanksref{e6, addr1}
\and Payam~Pakarha\thanksref{addr1}
\and Francesco~Piastra\thanksref{e8,addr1}
\and Julien~Wulf\thanksref{addr1}
}

\thankstext{e1}{e-mail: laura.baudis@uzh.ch}
\thankstext{e6}{e-mail: alexkish@physik.uzh.ch}
\thankstext{e8}{e-mail: fpiastra@physik.uzh.ch}

\institute{Department of Physics, University of Z\"{u}rich, Winterthurerstrasse 190, CH-8057, Z\"{u}rich, Switzerland \label{addr1}}
\date{}

\maketitle

\sloppy

\begin{abstract} 

A small-scale, two-phase (liquid/gas) xenon time projection chamber ({\sl{Xurich\,II}}) was designed, constructed and is under operation at the University of Z\"urich. Its main purpose is to investigate the microphysics of particle interactions in liquid xenon at energies below 50\,keV, which are relevant for rare event searches using xenon as target material. Here we describe in detail the detector, its associated infrastructure, and the signal identification algorithm developed for processing and analysing the data. We present the first characterisation of the new instrument with calibration data from an internal $^{83\mathrm{m}}$Kr source.
The zero-field light yield is 15.0 and 14.0 photoelectrons/keV at 9.4\,keV and 32.1\,keV, respectively, and the corresponding values at an electron drift field of 1\,kV/cm are 10.8 and 7.9 photoelectrons/keV. The charge yields at these energies are 28 and 31~electrons/keV, with the proportional scintillation yield of 24~photoelectrons per one electron extracted into the gas phase, and an electron lifetime of 200~$\mu$s. 
The relative energy resolution, $\sigma/E$, is  11.9\% and 5.8\%  at 9.4\,keV and 32.1\,keV, respectively using a linear combination of the scintillation and ionisation signals. We conclude with measurements of the electron drift velocity at various electric fields, and compare these to literature values.

\end{abstract}

\section{Introduction}
\label{sec:intro}

Radiation detectors using the noble gas xenon in its liquid form, with energy thresholds in the keV range, are of interest for direct dark matter detection experiments, searches for solar axions and axion-like particles, detection of low-energy solar neutrinos and measurements of coherent neutrino nucleus scattering, as well as for other rare event searches~\cite{Chepel:2012sj}.  In particular the sensitivity of direct dark matter searches has seen a dramatic increase over the last decade, a development which was largely lead by the dual-phase, xenon time projection chamber (TPC) technique~\cite{Baudis:2012ig,Undagoitia:2015gya,Baudis:2016qwx}. Large liquid xenon TPCs such as LUX~\cite{Akerib:2012ys}, PandaX~\cite{Cao:2014jsa} and XENON1T~\cite{Aprile:2017aty} constrain the cross section of weakly interacting massive particles  on nucleons down to 8$\times$10$^{\--47}$~cm$^2$~\cite{Akerib:2016vxi,Cui:2017nnn,Aprile:2017iyp}, while the upcoming XENONnT~\cite{Aprile:2015uzo}, LUX-ZEPLIN~\cite{Akerib:2015cja}, and the planned DARWIN detector~\cite{Aalbers:2016jon} are expected to improve upon these results by more than one and two orders of magnitude, respectively~\cite{Schumann:2015cpa}. The DARWIN detector will also search for the neutrinoless double beta decay of $^{136}$Xe, measure the low-energy solar neutrino flux with $<$1\% precision, observe coherent neutrino-nucleus interactions and detect galactic supernovae~\cite{Baudis:2013qla,Aalbers:2016jon,Lang:2016zhv}.

As part of our R\&D studies related to rare event searches, we have designed, constructed and are operating a new, small liquid xenon TPC \mbox{\it{(Xurich~II)}} at the University of Z\"urich. This builds upon our experience with larger TPCs, such as employed in XENON10/100/1T, and with a previous, small-scale TPC \mbox{\it{(Xurich~I)}}. The latter instrument demonstrated the capability of a spatially uniform calibration of liquid xenon detectors with $^{83\text{m}}$Kr~\cite{Manalaysay:2009yq} and was used to study the response of liquid xenon to electronic recoils down to 1.5\,keV~\cite{Baudis:2013cca}.

In a dual-phase (liquid-gas) TPC, the interactions of particles are observed via two distinct signals: the first is the prompt scintillation light (S1), while the second is caused by ionisation electrons that are drifted and extracted into the gaseous phase where they produce electroluminescence (S2). The photons are detected by photomultiplier tubes (PMTs) and the difference in arrival time between the S1 and S2 signals yields the depth, or $z$-position, of an interaction. The S2 light distribution in the PMTs yields the $(x,y)$-position of an interaction, while the S2/S1 ratio allows to distinguish between electronic recoils (ERs) and nuclear recoils (NRs)~\cite{Aprile:2009dv,Aprile:1900zz}.

This article is structured as follows: we introduce the new detector in section~\ref{sec:setup}, together with the description of its data acquisition and trigger system. In section~\ref{sec:analysis}, we detail the data processing and analysis of the S1 and S2 signals. We present the main results from calibration measurements in section~\ref{sec:results}, including our measurements of the electron drift velocity as a function of the applied electric field. In section~\ref{sec:summary} we summarise our results, and outline the near-future goals of the project.

\section{The Xurich\,II detector}
\label{sec:setup}

\subsection{Instrumentation}

\begin{figure}[h!]
\centering
\includegraphics*[width=0.35\textwidth]{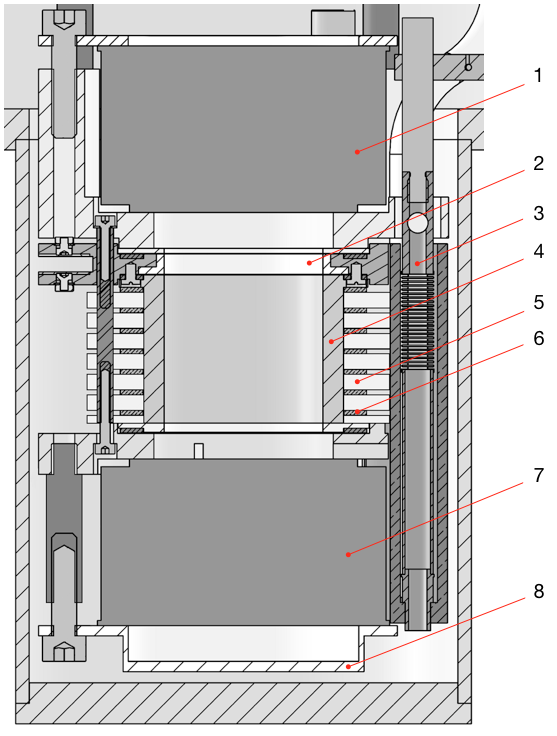}
\caption{\small {Schematic figure of the {\sl Xurich\,II} TPC. The liquid xenon (LXe) is contained within a structure made of PTFE and seven Cu field shaping rings. Two 2-inch PMTs view the active detector region, one placed in the liquid, and one in the gaseous phase on top. The electron drift and extraction fields are defined by three electrodes that are made of etched stainless steel meshes. Legend: 1 - top PMT, 2 - Torlon extraction spacer, 3 - liquid level control, 4 - inner PTFE reflector, 5 - PTFE drift spacers/insulators, 6 - copper field shaping rings, 7 - bottom PMT, 8 - PMT holder (PTFE).}}
\label{fig::tpc}
\end{figure}

The {\sl{Xurich\,II}} dual-phase TPC contains an active volume of 3.1\,cm diameter and 3.1\,cm height of xenon, for a total mass of 68\,g, assuming a liquid xenon density of 2.92\,g/cm$^{3}$. The structure delineating the cylindrical xenon volume is made out of polytetrafluoroethylene (PTFE), as well Polyamide-imide (Torlon) and polyetheretherketone (PEEK) for the non-reflective components. It is viewed by two circular, 2-inch photomultiplier tubes (PMTs), one placed in the liquid and one in the gaseous phase above the liquid, as shown schematically in figure~\ref{fig::tpc}. An electric field is defined by a set of three electrodes. The electron drift field is maintained between the cathode, at negative potential and located above the bottom PMT, and a grounded gate mesh, placed a few mm below the liquid xenon surface. The drift field uniformity is ensured by seven copper field shaping rings, separated by PTFE spacers. The stronger field, necessary to extract electrons into the vapour phase, is produced between the gate and an anode mesh that are held in place by a Torlon spacer. The anode is located a few mm above the liquid-gas interface. The liquid level is regulated by a weir system and a motion feedthrough, and must be accurately controlled to allow for two-phase operation. Three plate capacitors allow us to determine the level with 10\,$\mu$m precision, computed as an RMS of the baseline fluctuations. One 5\,cm long cylindrical capacitor is placed outside the active volume in order to monitor the liquid level during filling of the detector.

The PMTs are of type R9869 from Hamamatsu Photonics, developed for liquid xenon applications. They feature synthetic silica (SiO$_2$) windows and 12 stages of amplification. The quantum efficiency of the bialkali photocathode at 175\,nm is $\sim$35\%, the photocathode coverage is 16~cm$^{2}$.
The PMTs are operated in a negative bias configuration with the anode at ground, and the high-voltage (HV) is supplied by a CAEN N1470 module. The HV is distributed to the photocathode and the dynodes by voltage dividers (PMT bases) with a total resistance of 3.55\,M$\mathrm{\Omega}$. The heat dissipation is $\sim$100\,mW per base.

The TPC is contained within a stainless steel vessel, located inside a vacuum cryostat, with cooling provided by a copper cold finger immersed in a liquid nitrogen bath. A constant temperature is maintained by a 5\,W heater at the top flange of the inner cryostat vessel. The xenon is constantly purified by circulating it through a hot metal getter ({\sl{SAES MonoTorr}}). The gas handling and purification systems are described in~\cite{Manalaysay:2009yq}, where we employed a different TPC in an otherwise identical setting. The gas system is equipped with a small chamber that allows us to introduce the metastable $^{83\mathrm{m}}$Kr calibration source (T$_{1/2}$ = 1.83\,h) into the xenon gas flow, and thus into the TPC. The krypton source is produced by the decay of $^{83}$Rb (T$_{1/2}$ = 86.2\,d) embedded in zeolite, and is described in detail in~\cite{Manalaysay:2009yq}.

The achieved ionisation electron lifetime, which is a measure of the liquid xenon purity, is (198$\pm$8)~$\mu$s, corresponding to an electron mean-free-path of $\sim$40\,cm. This value is sufficient for our needs, because the maximum drift distance in the TPC is 3.1\,cm ($\sim$20~$\mu$s).

The performance of the detector is constantly monitored by a series of temperature, pressure and gas flow sensors. The temperatures inside the liquid xenon and at the top inner cryostat flange to which the heater is coupled, the pressure inside the TPC, the heater output power, and the gas recirculation flow are read out and and displayed on a website, where the values are updated once per minute. If any of the parameters exceeds a certain pre-defined range, email and SMS alarm messages are issued.

\subsection{Electric field configuration}

\begin{figure}[h!]
\centering
\includegraphics*[width=\columnwidth]{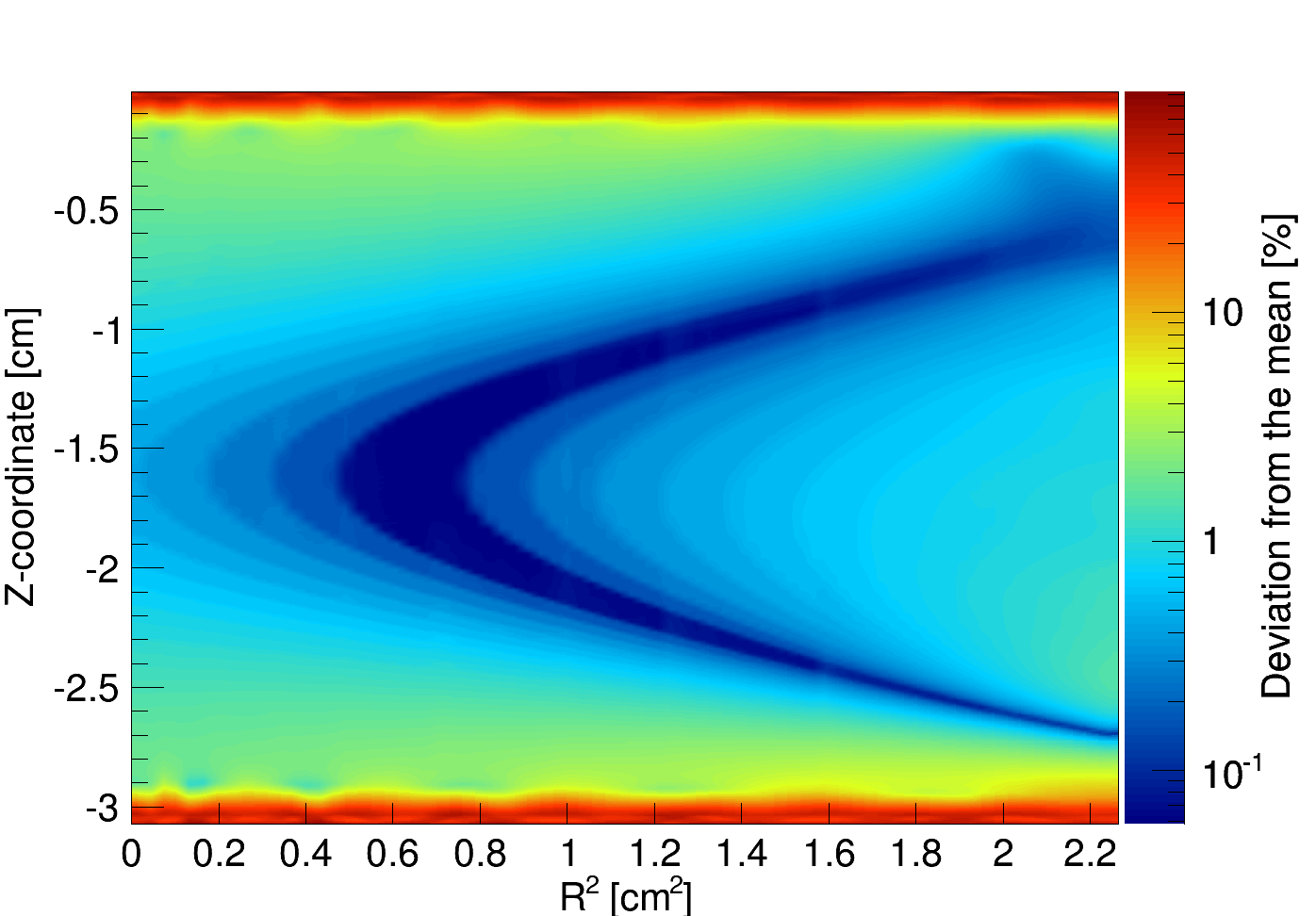}
\caption{\small {The electric field uniformity map in the detector volume, simulated with KEMfield~\cite{KEMfield}. The top and bottom of the plot correspond to the gate and cathode electrodes, respectively. The mean deviation from field uniformity in the target volume is 2.8\%. The nominal fiducial volume cut used for analysis removes 3\,mm from the top and bottom, reducing the mean field deviation from uniformity to 0.9\%.}}
\label{fig:ElectricField2D}
\end{figure}

The design of the electric field cage of the TPC was optimised based on simulations with COMSOL~\cite{HrvojeThesis} and KEMfield~\cite{KEMfield}, with the goal of maximising the field uniformity and the optical transparency of the electrodes. The resulting, two-dimensional field map is shown in figure~\ref{fig:ElectricField2D}. 
The electrodes (cathode, anode and gate) are made out of chemically etched stainless steel meshes with thickness and wire diameter of 100\,$\mu$m and pitch of 2.7\,mm, respectively, resulting in 93\% geometrical optical transparency.

Since only 2 non-segmented PMTs are employed to detect the scintillation light in the TPC, the reconstruction of the $(x,y)$ interaction vertex is not possible. Hence the fiducialisation of the target volume is performed with the $z$-coordinate of an interaction only, which is reconstructed with a resolution of 0.2\,mm (1$\sigma$). The deviation of the electric drift field from uniformity in the target volume is 2.8\%. Fiducialising the target by removing 3\,mm from the top and bottom of the liquid removes high-field regions close to the cathode and gate electrodes and reduces the non-uniformity with respect to the volume-averaged value to 0.9\%.

The field cage and the dedicated voltage divider circuit with a total resistivity of 1.05\,G$\mathrm{\Omega}$ allow us to apply a high voltage to the cathode of up to 6\,kV. The nominal potential on the anode is 4\,kV, for an extraction field of (10.32$\pm$0.14)~kV/cm, resulting in complete extraction of ionisation electrons from the liquid to gas phase~\cite{Aprile_ExtractionEfficiency,Xe100_SingleElectrons}.

\subsection{The data acquisition system}

The data acquisition (DAQ) system generates the trigger, digitises the waveforms of the two PMTs, and stores the data to disk. The signals from the PMT bases are digitised by a CAEN V1724 Flash ADC with 10\,ns sampling period, 2.25\,V full scale, 14-bit resolution and 40\,MHz bandwidth, after passing through a CAEN 625 fan-in/fan-out module.

For the trigger, generated by a leading edge discriminator (CAEN N840), we nominally require that the top PMT signal height exceeds a threshold of 10~mV, which corresponds to a deposited energy of 0.6~keV. The trigger acceptance has been measured with a pulse generator and is 100\% above 8~mV. Regardless of whether the trigger is generated by an S1 or an S2 signal, it is placed in the middle of the event window of 60~$\mu$s width, sufficiently larger than the maximum electron drift time of 19~$\mu$s at a field of 220~V/cm. The event rate during runs with the $^{83\mathrm{m}}$Kr calibration source does not exceed $\sim$100~Hz, resulting in a pile-up fraction below 0.6\%.

\section{Signal processing}
\label{sec:analysis}

The signal processing starts from the pulse identification and calculation of the relevant peak quantities, and is followed by conversion of the measured PMT charge into the unit of photoelectrons (PE). Subsequently, a correction to the scintillation signal is applied to account for spatial variations in the light collection efficiency.

\subsection{Pulse identification}

The identification of S1- and S2-like pulses is performed with an algorithm which employs two different width-based filters and an additional $\chi^{2}$-filter. Prior to applying the filters, the signal baseline height and its RMS noise amplitude are quantified using the first and last 50 time samples (1 sample = 10\,ns) of each digitised waveform, on an event-by-event basis, and subtracted from the signal trace. Events where the two baseline values differ by $>$3$\times$RMS are discarded from further analysis.

The width-based filters are defined as:
\begin{align}
~~~~~~~A_{i}[1] &= \sum^{i+\frac{w_1}{2}}_{j=i-\frac{w_1}{2}}~S_{j}, \label{eq:S1filter}\\
~~~~~~~A_{i}[2] &= \sum^{i+\frac{w_2}{2}}_{j=i-\frac{w_2}{2}}~S_{j} - \max_{j \in [i-\frac{w_2}{2}, i+\frac{w_2}{2}]} A_{j}[1],\label{eq:S2filter}
\end{align}

\noindent
where $S_{j}$ is the amplitude of the baseline-subtracted waveform corresponding to bin $j$, and $w_1$ and $w_2$ are boundary conditions for the S1 and S2 signal widths, respectively. Due to the relatively short decay constants of the xenon scintillation light, (4.3$\pm$0.6)\,ns and (22.0$\pm$1.5)\,ns for the singlet and triplet components, respectively \cite{Doke:1999ku}, and a fast transit time of the PMTs  ($<$~20\,ns \cite{XurichPMT}), all information about the S1 pulse shape is contained within a time window $w_1$\,=\,80\,ns. The S2 signal full width at tenth maximum (FWTM) does not exceed 0.8\,$\mu$s, thus $w_2$\,=\,1.1\,$\mu$s is a sufficient time interval to contain the entire pulse. Because of the width of the S2 signal (FWHM$<$0.35\,$\mu$s), the efficiency to separate two S2 pulses is unity if the corresponding particle interactions occur at a distance $>$1\,mm apart in \textit{z}.

\begin{figure}[h!]
\centering
\includegraphics*[width=\columnwidth]{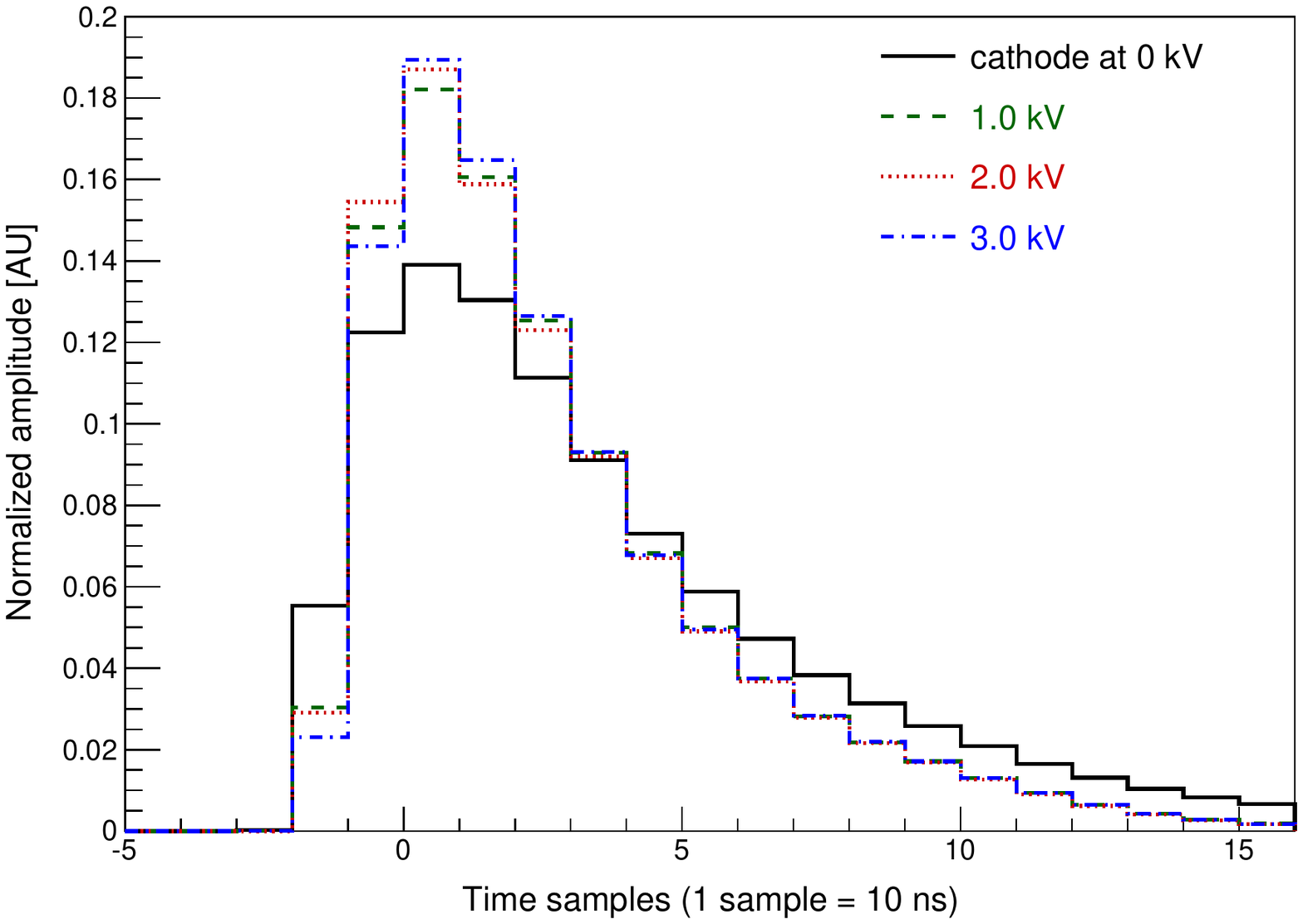}
\caption{\small {The S1 signal templates for the bottom PMT, derived from calibration data acquired with the grounded anode mesh (hence no S2 signals present) and various electric drift field settings. These range from 0\,kV/cm (cathode at ground, black solid) to 1\,kV/cm (cathode at 3\,kV, blue, dot-dashed). The $\chi^{2}$ filter employed in the data processing is calculated using the S1 template defined at the same electric field. The histograms are normalised by their area.}}
\label{fig:S1template}
\end{figure}

The pulse identification algorithm first scans each signal trace to look for S2 candidates within a search region defined by $A_{i}[2]>0$. In each trace where at least one S2 candidate is identified, the filter is applied to detect S1 pulses in a search region where $A_{i}[1]>0$. Inside these regions a candidate S2 is selected if $A_{i}[2]>A_{i}[1]$ in every time sample $i$; otherwise the entire trace is removed from the S2 search.
After this first operation, a $\chi^{2}$-filter is applied to the selected pulse candidates. The $\chi^{2}$-filter is based on a template calculated for each PMT as a function of applied drift field, as shown in figure~\ref{fig:S1template}. The templates are built from $\sim$10$^{4}$ S1 signals acquired with the extraction field (anode potential) set to zero, hence no electroluminescence (S2) signals are generated. 
The selected S1 signals are aligned on the time sample corresponding to their maximum amplitude, then the amplitude in each time sample is computed as the median of all accumulated signals.
The resulting signal shape is normalised by its total area.

Within each waveform, the similarity of a candidate pulse with an S1-like signal is measured by the value of the $\chi^{2}$-filter, defined as:
\begin{equation}
\label{eq:Chi2filter}
~~~~~~~\chi^{2}_{i}= \sum_{k=-l}^{r} \left( T_{k} - \tilde{S}_{i+k} \right)^{2},
\end{equation}

\begin{figure}[h!]
\centering
\includegraphics*[width=\columnwidth]{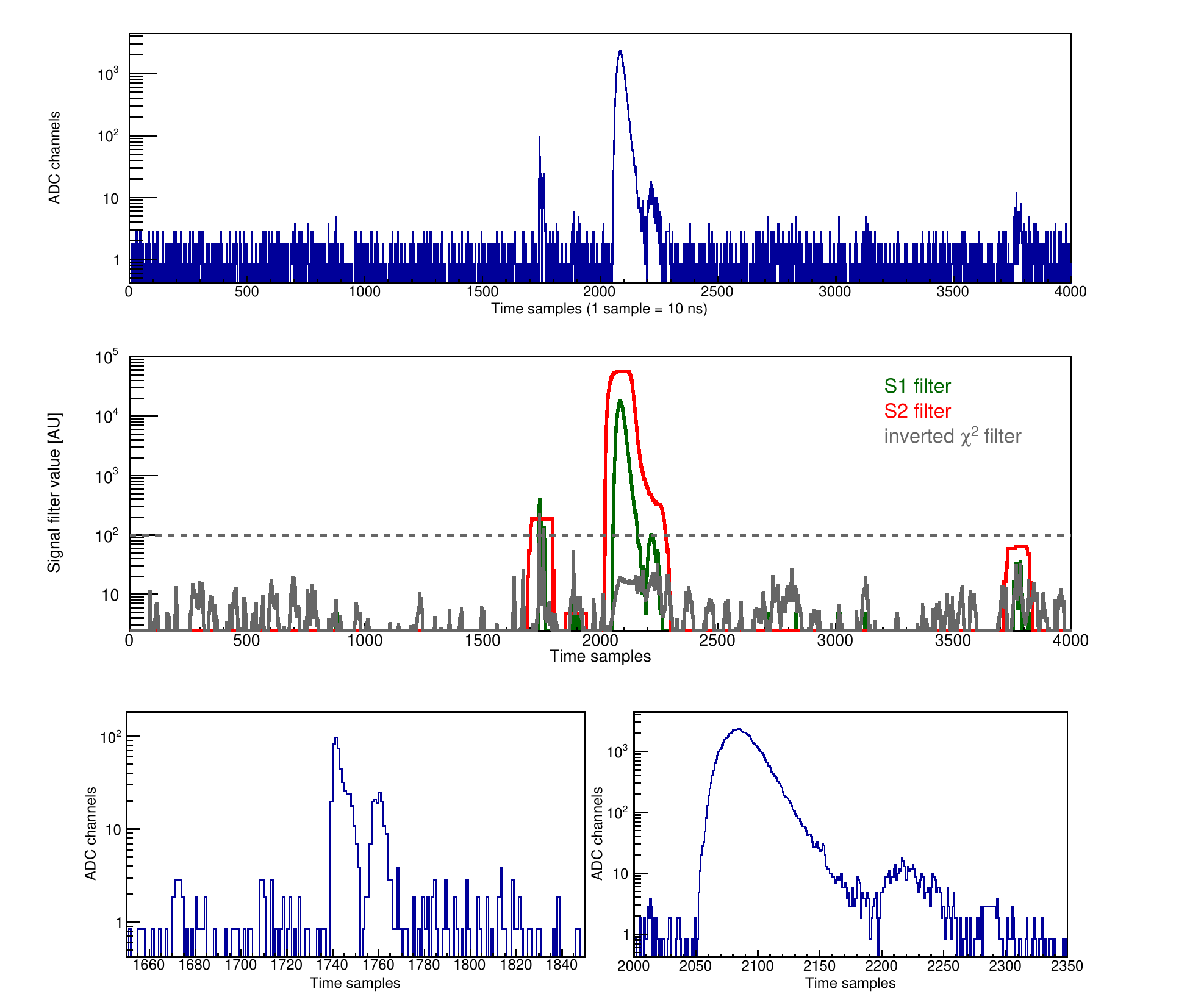}
\caption{\small A signal trace acquired with the bottom PMT from calibration with the $^{83\mathrm{m}}$Kr source, top. The middle panel illustrates the performance of the pulse identification algorithm, where width-based S1 and S2 filters are shown in green and red, respectively. The value of the inverted $\chi^2$ filter is shown in grey, with the threshold represented by the dashed grey line. The bottom panel shows a zoom into the regions where S1 and S2 pulses were identified.}
\label{fig:SignalTrace}
\end{figure}

\noindent
where $T_{k}$ is the template amplitude in sample $k$, and $l$ and $r$ are the numbers of samples to the left and right of the sample that corresponds to the template maximum ($k=0$). The $\tilde{S}_{i}$ is the amplitude of the waveform at time sample $i$, normalised by the trace area between sample ($i-l$) and sample ($i+r$), which corresponds to the interval overlapping with the template.

Within each waveform, a candidate S2 is accepted if the value of the $\chi^{2}$-filter for the time sample that corresponds to its maximum is higher than a set threshold. 
When an S2 pulse is accepted, its properties are quantified using all time samples to the right and to the left of the maximum, until the S2 filter reaches either zero or a relative minimum. The latter condition indicates the presence of an additional S2 in the signal trace; in this case the filter is applied by the same method to the remaining part of the search region. If, on the contrary, the value of the $\chi^{2}$-based filter is lower than the threshold, the pulse is identified as an S1 candidate. 
Once all S2 search regions in the waveform have been evaluated, the S2 peak finder loop terminates, and the S1 peak finder iterates over all the S1 search regions to evaluate the candidate pulses.
For each selected S1 signal the respective pulse properties are computed using the same criteria as for the S2 signals. The S1 searching loop ends when all S1 regions in the waveform have been evaluated.

The performance of the pulse identification algorithm is illustrated in figure~\ref{fig:SignalTrace} on an example of a signal trace acquired with the $^{83\mathrm{m}}$Kr source. From the study of selected data sets processed with and without the additional $\chi^{2}$-filter, we concluded that this auxiliary tool for signal identification efficiently detects small S1-like pulses very close to S2 signals, usually corresponding to interactions located in the 2\,mm region between the gate mesh and the liquid surface. In addition, the $\chi^{2}$-filter is able to identify S1s with the size of a few photoelectrons surrounded by noise and initially misidentified as S2 signals.

\subsection{Photomultiplier gain calibration}

\begin{figure}[h!]
\centering
\includegraphics*[width=\columnwidth]{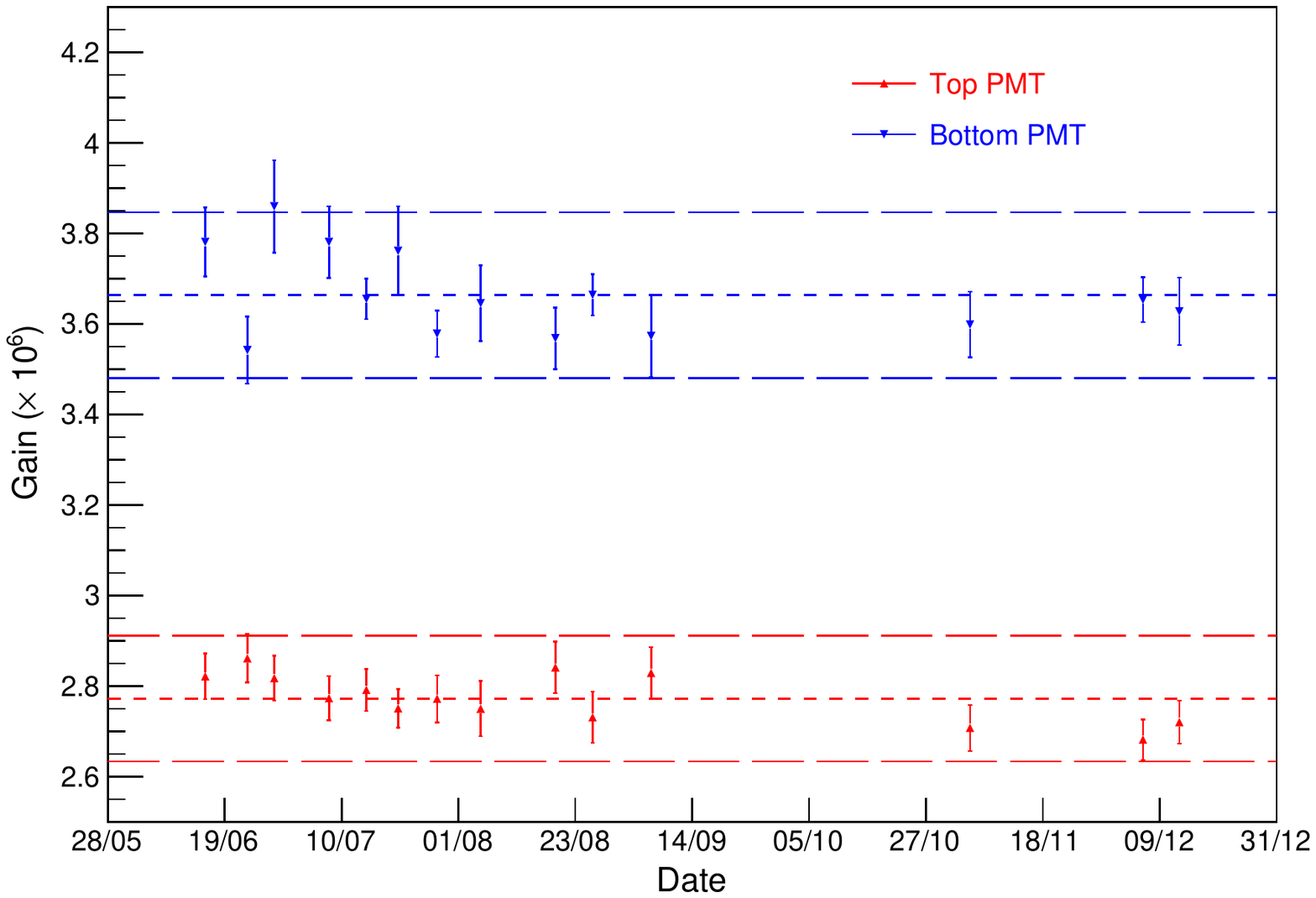}
\caption{\small {The gains of the top (black) and bottom (red) PMTs, shown here over a period of six months, are stable within 3\%.  The horizontal lines indicate the mean and RMS spread of the measured gain values.}}
\label{fig:PMTgains}
\end{figure}

The gains of the PMTs are regularly calibrated with blue light ($\sim$470\,nm) from an external LED, transferred into the TPC via polymethylmethacrylate (PMMA) optical fibres. We use two methods to determine the PMT gain. In the first, fit-based method, the light level is adjusted such that a PMT shows a signal in the time window considered for analysis in $\sim$5\% of all triggers. The second approach is based on a model-independent method described in~\cite{Saldanha:2016}, without making any assumptions about the underlying single photoelectron distribution. Both methods yield consistent results: the gains are $(2.90\pm0.04)\times10^6$ at 870\,V for the top, and $(3.73\pm0.09)\times10^6$ at 940\,V for the bottom PMT. These have been chosen to avoid non-linear effects in the PMTs and electronics for calibration runs. The quoted uncertainty is the RMS of the gain measurements performed regularly over a period of six months, showing that the values are stable within 3\% (see figure~\ref{fig:PMTgains}). The peak areas are converted into photoelectrons using the time-averaged gain value for each PMT.

\subsection{Light collection efficiency}

Geometrical and optical properties of the TPC lead to non-uniformities in the scintillation light collection efficiency (LCE) within the target volume. Therefore, a dedicated correction function is applied to compensate for non-uniformities in the detector response. Due to the absence of $(x,y)$-position reconstruction, only the depth-dependence can be studied and taken into account. 

\begin{figure}[b!]
\centering
\includegraphics*[width=\columnwidth]{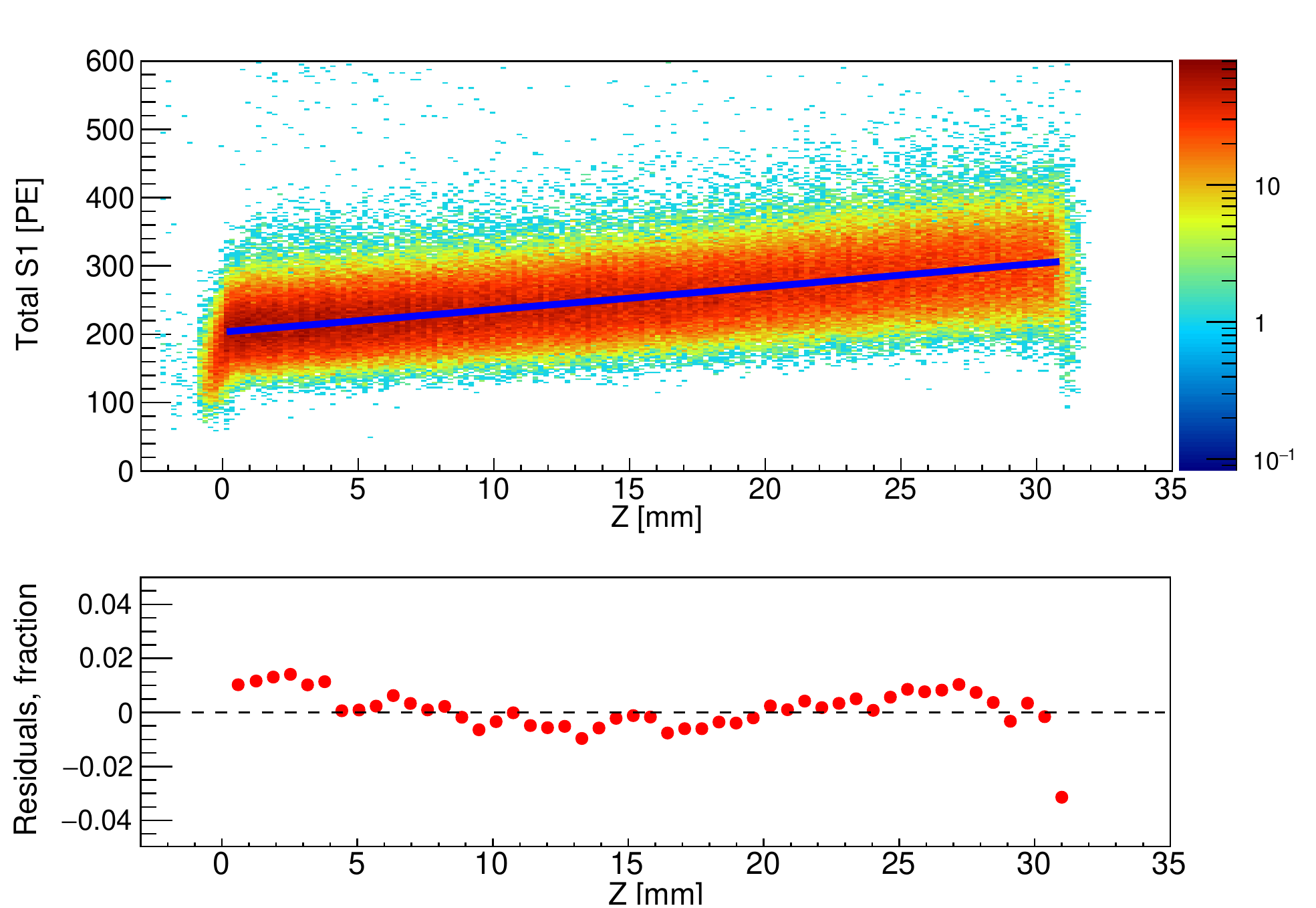}
\caption{\small {Dependency of the S1 signal on the electron drift time, and hence on the depth $z$ of the interaction, measured with a $^{83\mathrm{m}}$Kr calibration source. The correction function, a 2nd-order polynomial fit, is represented by the blue line (top) with residuals from the fit $<$2$\%$ (bottom).}}
\label{fig:S1correction}
\end{figure}

A spatial correction map for the scintillation signal from each PMT and their sum has been determined {\sl in-situ} using $^{83\mathrm{m}}$Kr data. Figure~\ref{fig:S1correction}, top, shows the summed signals from the 32\,keV line. From this data one can see the S1 dependence as a function of interaction depth. A correction function was obtained by first performing a Gaussian fit on the S1 distribution of each slice ($\pm$0.5\,mm around each $z$-position). A second order polynomial approximation was fit to the mean of each slice in a pre-defined analysis volume, -25\,mm $< z <$ -5\,mm, and used to analytically obtain the corrected S1 signals. The residuals from the fit are shown in figure~\ref{fig:S1correction}, bottom. The variation in S1 after applying the correction in this fiducial region is $<$2\%.

To benchmark the position-dependent LCE for the S1 signals, Monte Carlo simulations have been performed and the results compared with measured data. From 10$^{6}$ interaction vertices within the simulated volume, 10$^{3}$ photons of wavelength 175\,nm were generated isotropically and with random polarisation. The detector model takes into account relevant physical processes, e.g. light absorption, reflection and refraction at the surface boundaries of each material, Rayleigh scattering, attenuation and transport of the photons. The electrodes were encoded to allow for 93$\%$ transparency and a refractive index to match the one of gaseous xenon (anode) and LXe (gate, cathode). Additionally, the refractive index of the PMT window was encoded, and the photocathodes were modelled for unitary QE, allowing for full absorption of the propagated photons. The most relevant optical parameters are listed in Table~\ref{table:XuOptParams}.

\begin{table}[ht]
\centering
\begin{tabular}{ll}

{Parameter} & {Value} \\  
\midrule
LXe refractive index & 1.63\tabularnewline
LXe Rayleigh scattering length & $30\,$cm\tabularnewline
LXe absorption length & $50\,$m\tabularnewline
Gas Xe refractive index & 1.0\tabularnewline
Gas Xe Rayleigh scattering length & $100\,$m\tabularnewline
Gas Xe absorption length & $100\,$m\tabularnewline
PTFE refractive index & $1.58$\tabularnewline
PTFE reflectivity & $0.95$\tabularnewline
\end{tabular}
\caption{\small{The most relevant optical parameters for simulations of the S1 light propagation and LCE predictions, as encoded in the Geant4 model.}}
\label{table:XuOptParams}
\end{table}

Primarily due to internal reflection at the liquid-gas interface from the higher refractive index of LXe as compared to GXe, it was observed in both simulation and data that most of the S1 light is collected by the bottom PMT. The unbinned spatial LCE distributions were fit to obtain mean LCE values for top, bottom, and both PMTs, yielding $(12.5\pm0.1)\%$, $(47.0\pm0.1)\%$, and $(59.8\pm0.1)\%$, respectively. The radial dependence was observed to be highly uniform in the simulations, with a slight decrease in light collection at larger radii. The radial variations relative to the mean LCE were $\pm5\%$, $\pm0.5\%$, and $\pm1.5\%$ for top, bottom, and both PMTs, respectively. 

\begin{figure}[h!]
\centering
\includegraphics*[width=\columnwidth]{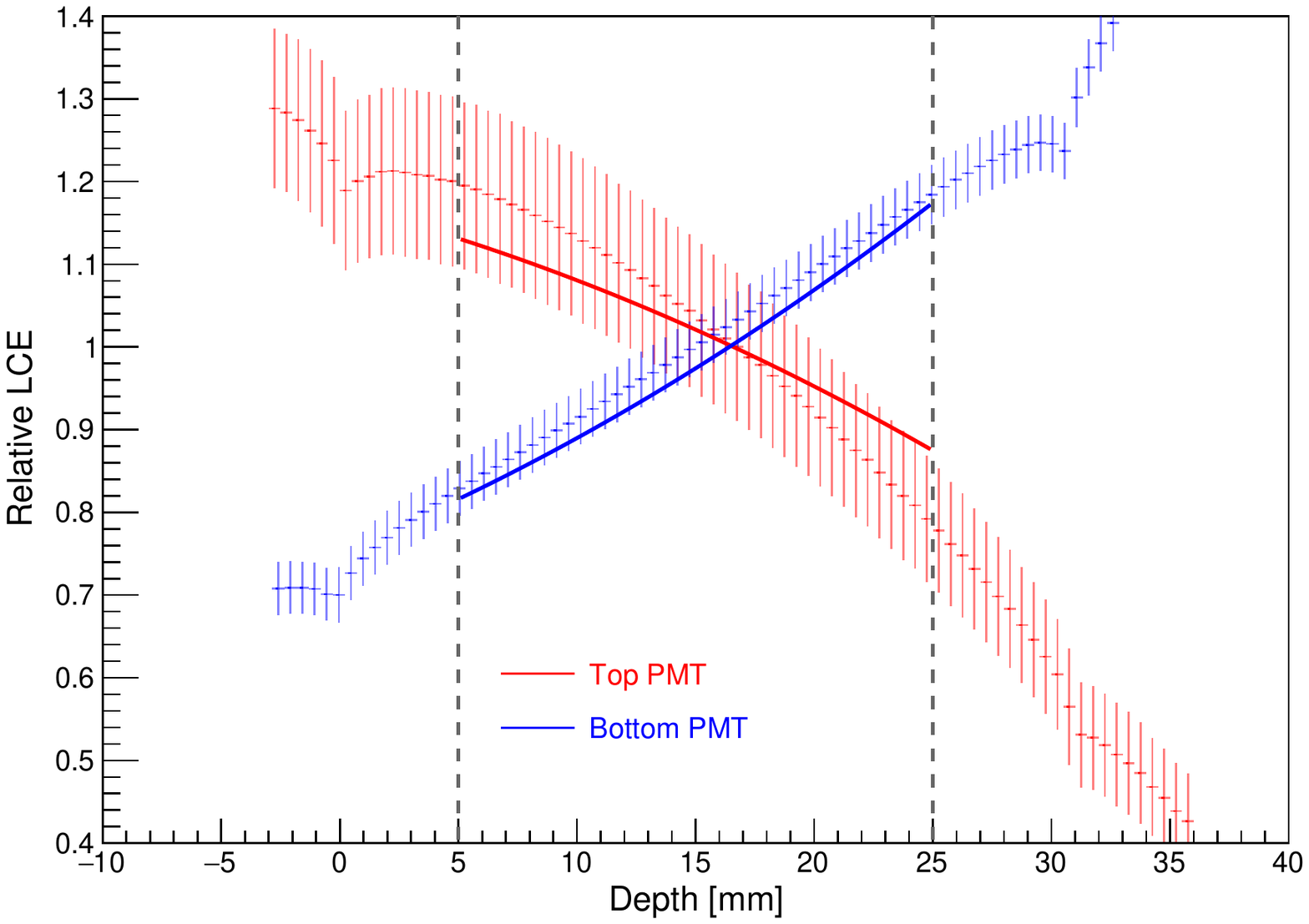}
\caption{\small {The mean LCE relative to the average over the fiducialised volume, indicated by the vertical dashed lines, as a function of interaction depth. A comparison of relative LCE is made between simulation (discrete points with one standard deviation error bars) and data after parametrisation (solid lines) for the top and bottom PMTs (red and blue, respectively).}}
\label{fig:RelativeLCE_v_Depth}
\end{figure}

The relative LCE, i.e. the mean light collection efficiency relative to the volume-averaged values given above, is shown in figure~\ref{fig:RelativeLCE_v_Depth} as a function of interaction depth using slices in $z$ of 0.5\,mm. We observe a good agreement between simulation (discrete points with error bars given by one sigma of each Gaussian fit slice) and measurement (solid lines) for top and bottom PMTs (red and blue, respectively) within the analysis volume as indicated by the vertical dashed lines. The match is within the $\pm1~\sigma$ predictions, thus verifying the results of the electron drift velocity measurement (see section~\ref{sec:EDV}) as well as the reconstruction of interaction depth. For the bottom PMT, the prediction slightly overestimates the LCE. The simulation also shows a much steeper decrease of the relative LCE as a function of depth with respect to the top PMT. For the latter, the photocathode absorption as a function of photon incidence angle may play a role. Further details can be found in~\cite{FrancescoThesis}. 

\section{Results}
\label{sec:results}

In this section we present first results obtained with the {\sl Xurich\,II} TPC. We first discuss the energy calibration and energy resolution of the detector, after which we show measurements of the electron drift velocity as a function of electric field and compare these to literature values.

\subsection{Energy calibration}

\begin{figure}[h!]
\centering
\includegraphics*[width=\columnwidth]{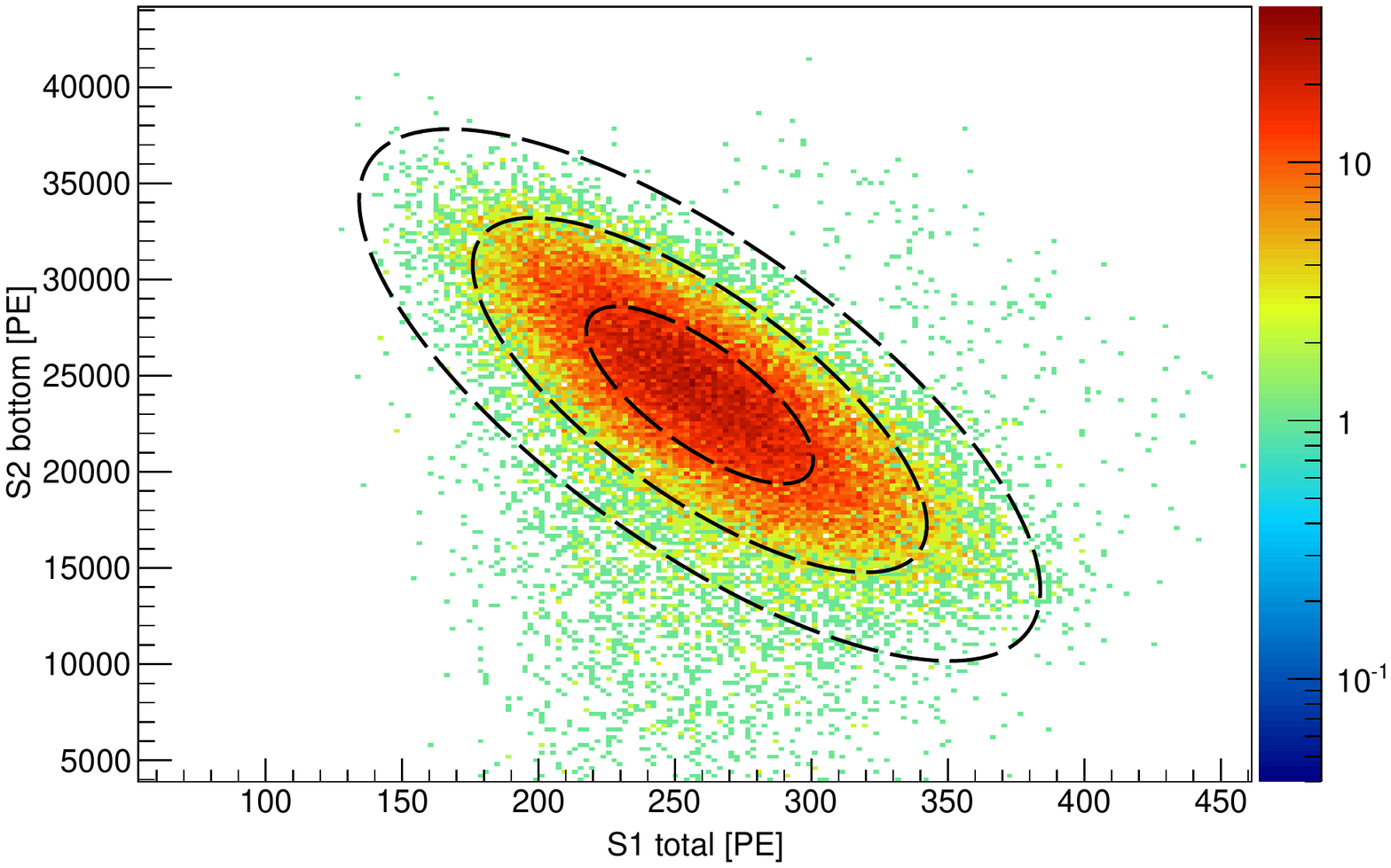}
\caption{\small Charge and light anti-correlation for the 32.1\,keV energy line from a calibration with $^{83\mathrm{m}}$Kr. The data were acquired at nominal TPC settings: cathode at 3\,kV, anode at 4\,kV. The S1 signal is corrected using the function shown in figure~\ref{fig:S1correction}. Events with lower S2 values are attributed to events close to the walls of the TPC, and thus reduced charge collection.}
\label{fig:calib_kr83m_s1}
\end{figure}

The energy calibration is performed with $^{83\mathrm{m}}$Kr, providing low-energy lines at 9.4\,keV and 32.1\,keV uniformly distributed within the target volume~\cite{Manalaysay:2009yq}. These are tagged by exploiting their double-S1 and double-S2 topology, given the measured half-life of the first excited state at 9.4\,keV of (155$\pm$1)\,ns. In figure~\ref{fig:calib_kr83m_s1} we show the anti-correlation between the scintillation and ionisation signals for the 32.1\,keV line with data from one calibration run.

\begin{figure}[h!]
\centering
\includegraphics*[width=\columnwidth]{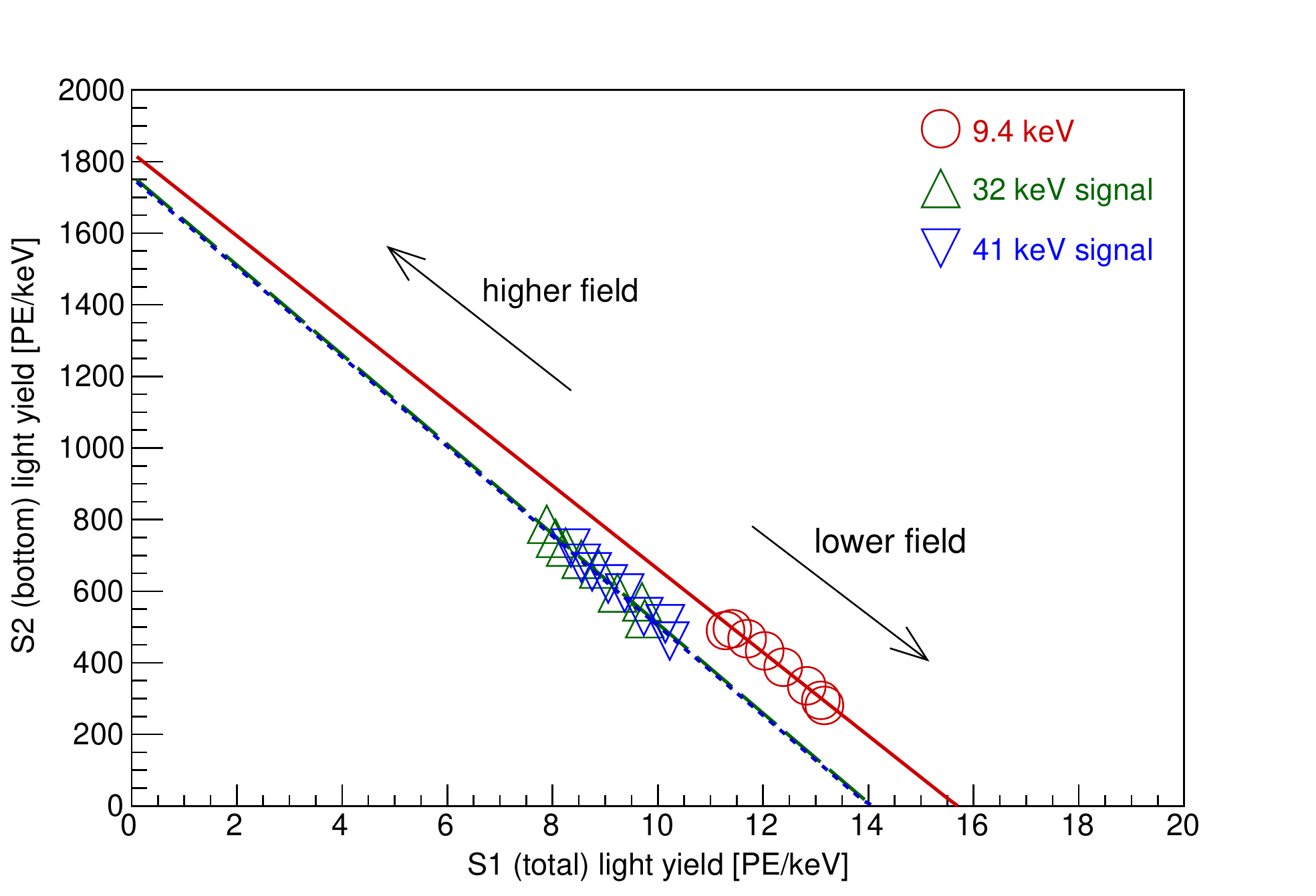}
\caption{\small Anti-correlation between scintillation and ionisation signals measured at drift fields in the range $(0.2-1.3)$\,kV/cm. The 9.4\,keV line yields a higher total number of quanta than the 32.1\,keV and 41.5\,keV lines, and we refer to the main text for an explanation.}
\label{fig:calib_kr83m_s2}
\end{figure}

By performing these measurements at various drift fields, we can observe the anti-correlation between scintillation light and ionisation in the TPC, as illustrated by the S1 and S2 photon yields for three energies in figure~\ref{fig:calib_kr83m_s2}. As expected, the proportion of light and charge changes at different drift fields, but their sum remains constant. The data points for 32.1\,keV and 41.5\,keV, the sum of 9.4\,keV and 32.1\,keV transition energies due to the short half-life, fall on the same line; this is expected, for the number of quanta is proportional to the deposited energy. However, this agreement is not observed for the 9.4\,keV signal, resulting in a higher apparent yield of quanta. Its origin is most likely the presence of the trailing tail of the 32.1\,keV S2 pulse underneath the following S2 signal, and the presence of spurious extracted electrons, originating from photoionisation of the gate electrode and of LXe impurities following the rather high intensity 32.1\,keV proportional scintillation light.

The detector-specific gains, $g_1$ and $g_2$, in units of detected photoelectrons/quantum, can be determined from the measured S1 and S2 signals, and are defined as:

\begin{equation}
E_{\rm{CES}} = W (n_{\gamma} + n_e) = W \left( \frac{S_1}{g_1} + \frac{S_2}{g_2} \right), 
\label{eq:ces}
\end{equation}

\noindent
where  $W$ = (13.7$\pm$0.2)~eV~\cite{WorkFunction} is the energy required to produce an excited or ionised xenon atom, $n_{\gamma}$ is the number of photons emitted by the excimer de-excitation and electron-ion pair recombination processes, $n_e$ is the number of electrons that escape recombination, and $E_{\rm{CES}}$ is the so-called combined energy scale.

\begin{figure}[h!]
\centering
\includegraphics*[width=\columnwidth]{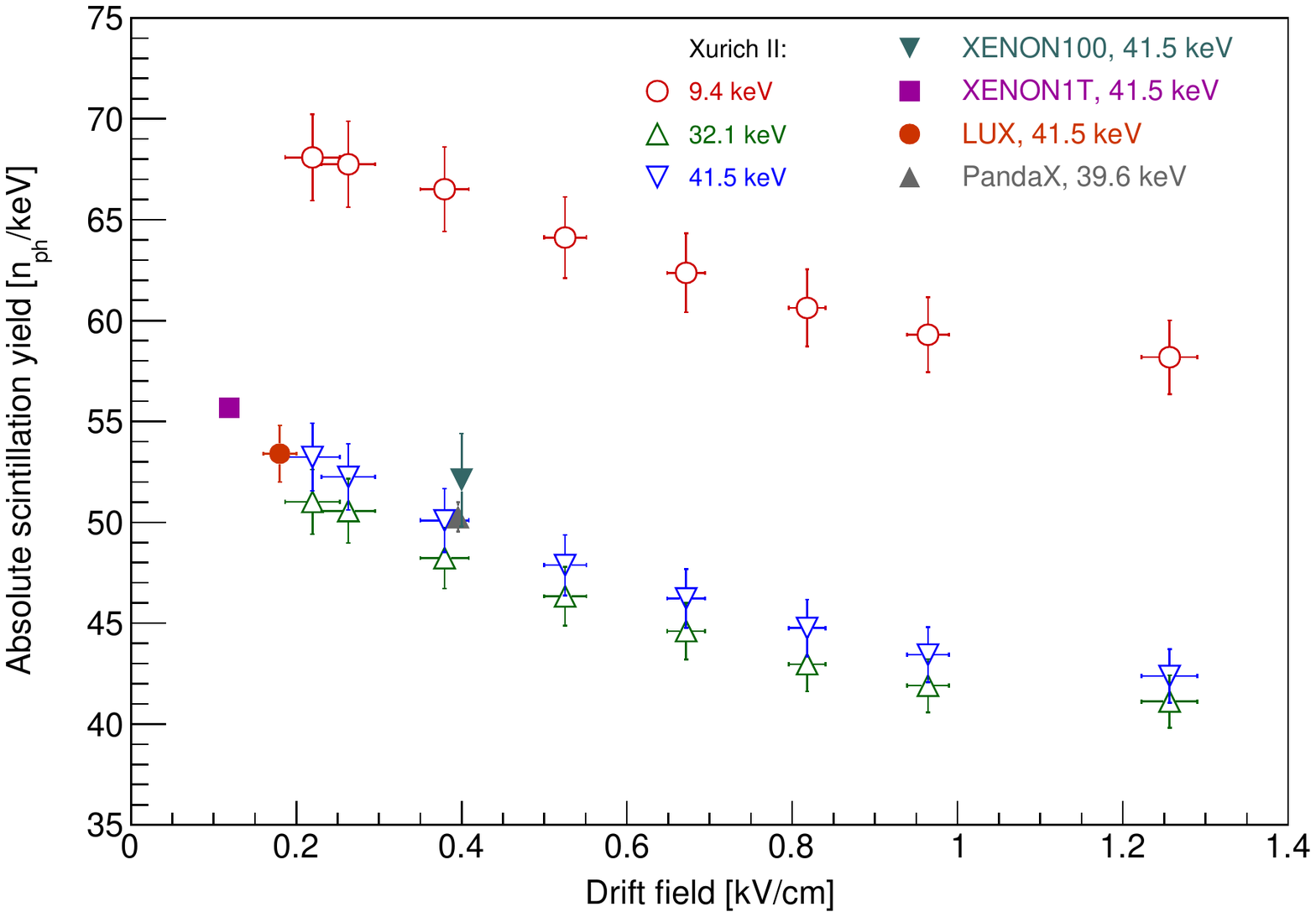}
\includegraphics*[width=\columnwidth]{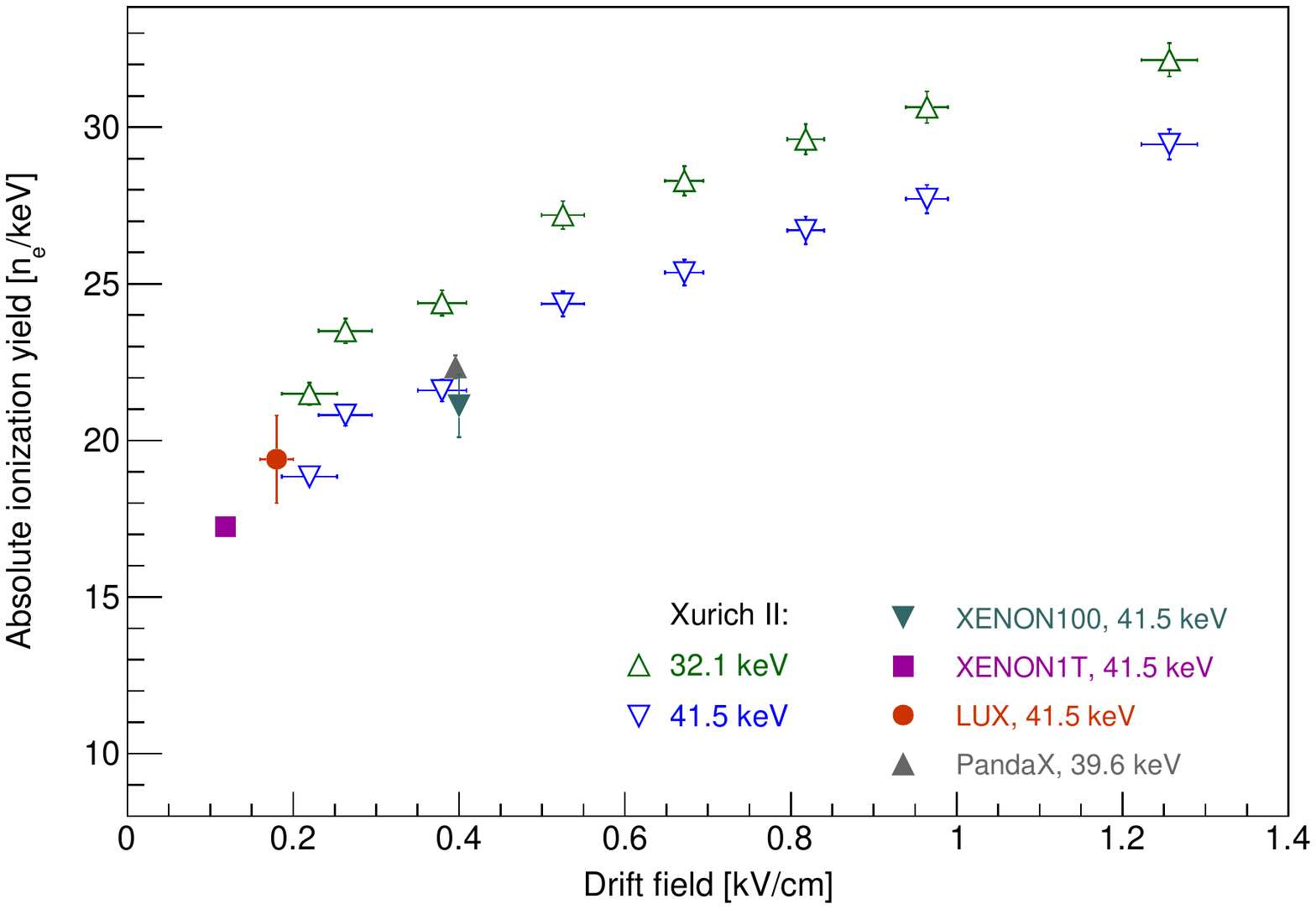}
\caption{\small Absolute scintillation ($top$) and ionisation ($bottom$) yields for the $^{83\mathrm{m}}$Kr energy calibration lines as a function of the electron drift field. The ionisation yield for 9.4\,keV transition is not presented due to systematic effects described in the text. Also shown are data points from LUX~\cite{Akerib:2015rjg}, XENON100~\cite{Aprile:2017xxh}, XENON1T~\cite{Aprile:2017aty}, and PandaX~\cite{Tan:2016zwf}.}
\label{fig:AbsYields}
\end{figure}

We determined the gains from the electric field dependency of S1 and S2, namely from the $x$- and $y$-intercepts in figure~\ref{fig:calib_kr83m_s2}.  The photon detection efficiency for prompt scintillation is $g_{1}$ = (0.191$\pm$0.006)\,PE/photon, affected by the light collection efficiency in the liquid xenon target and the quantum efficiency of the photomultiplier tubes. The charge amplification gain is $g_{2}$ = (24.4$\pm$0.4)\,PE/electron.  Combining these quantities with the light and charge yields for the $^{83\mathrm{m}}$Kr lines, a calibration of the prompt and proportional scintillation signals in terms of the absolute number of quanta is performed. This is shown as a function of the electron drift field in figure~\ref{fig:AbsYields}, together with the values obtained at low fields and at 41.5\,keV by large-scale TPCs such as XENON100~\cite{Aprile:2017xxh}, LUX~\cite{Akerib:2015rjg}, PandaX~\cite{Tan:2016zwf} and XENON1T~\cite{Aprile:2017aty}. The charge yield at 9.4\,keV is not shown due to systematic effects described above. The agreement between our results and the values from LUX, PandaX, XENON100, and XENON1T is remarkable, while our measurements extend to higher electric drift fields.

The absolute calibration is used to predict the potential of our detector to observe low-energy nuclear recoils, where the energy threshold is estimated based on the predictions of the NEST model~\cite{Szydagis:2011tk,Szydagis:2013sih}. An analysis threshold of 2\,PE corresponds to a mean number of 10.5 primary photons, which translates to an energy threshold of (2.3$-$2.7)\,keV nuclear recoil energy, depending on the drift field.

\begin{figure}[h!]
\centering
\includegraphics*[width=\columnwidth]{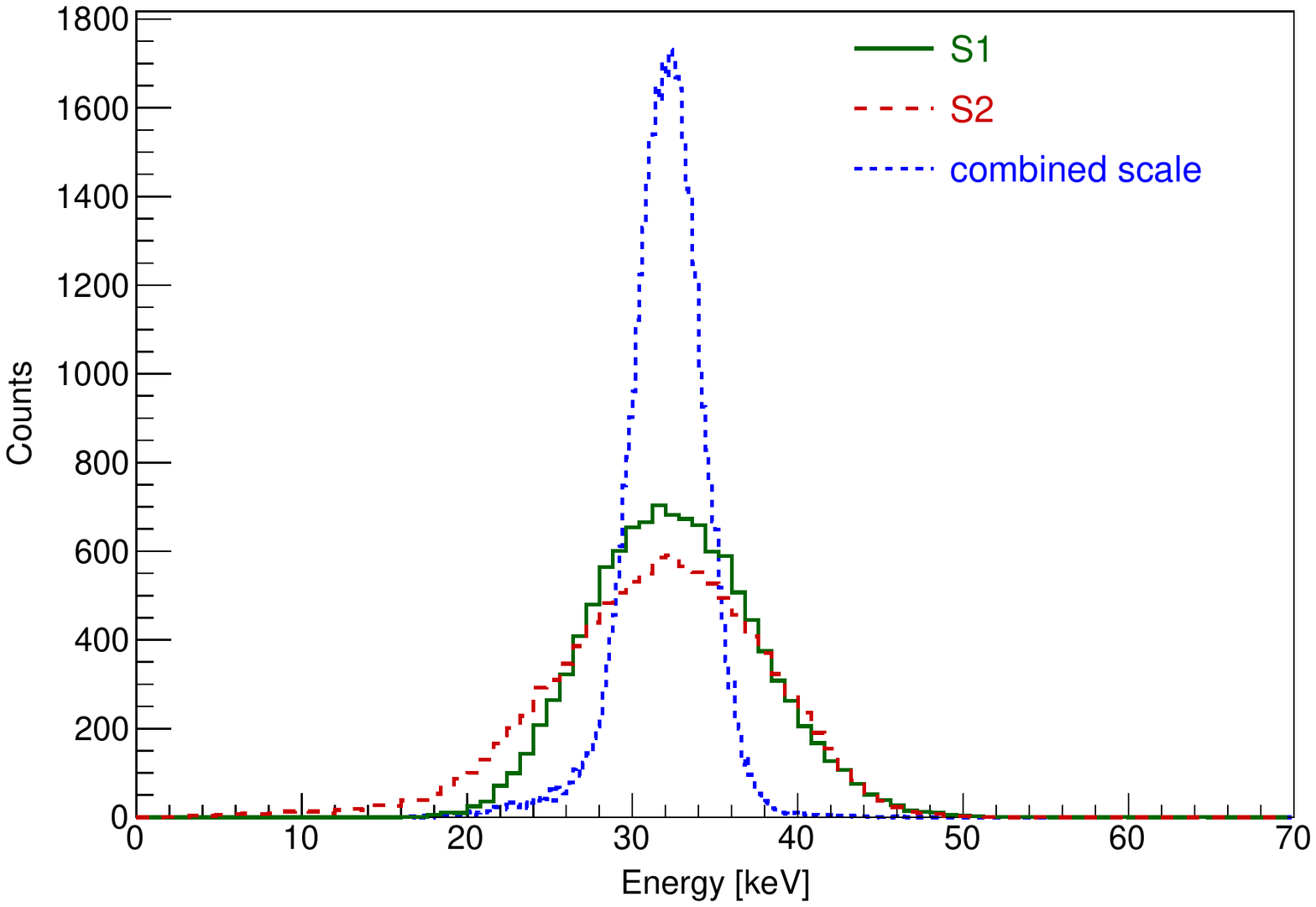}
\caption{\small Observed 32.1\,keV line from a calibration with the $^{83\mathrm{m}}$Kr source at a drift field of 1\,kV/cm. The spectrum is shown for different energy determinations, using S1 (green), S2 (red) and a linear combination of the two, $E_{\rm{CES}}$ (blue). The relative energy resolution ($\sigma$/E) is ($5.8\pm0.3$)\% for the combined energy scale.}
\label{fig:calib_kr83m_s3}
\end{figure}

The charge-light anti-correlation can be quantified and used to build an energy scale by combining both signals (see equation~\ref{eq:ces}), and thus to improve the energy resolution of the detector. The energy spectrum for the 32.1\,keV line is shown in figure~\ref{fig:calib_kr83m_s3}, together with the spectra reconstructed with the scintillation or ionisation signals alone. The energy resolution ($\sigma$/E) is ($16.6\pm0.1$)\% for S1, ($20\pm1$)\% for S2, and ($5.8\pm0.3$)\% for the combined energy scale. It is worth mentioning that the energy resolution for the S2-only case is generally expected to be superior compared to the S1-only case. This is not observed here, likely due to the absence of an $(x,y)$-position reconstruction and hence the presence of events close to the PTFE wall of the TPC with reduced charge collection.

\begin{figure}[h!]
\centering
\includegraphics*[width=\columnwidth]{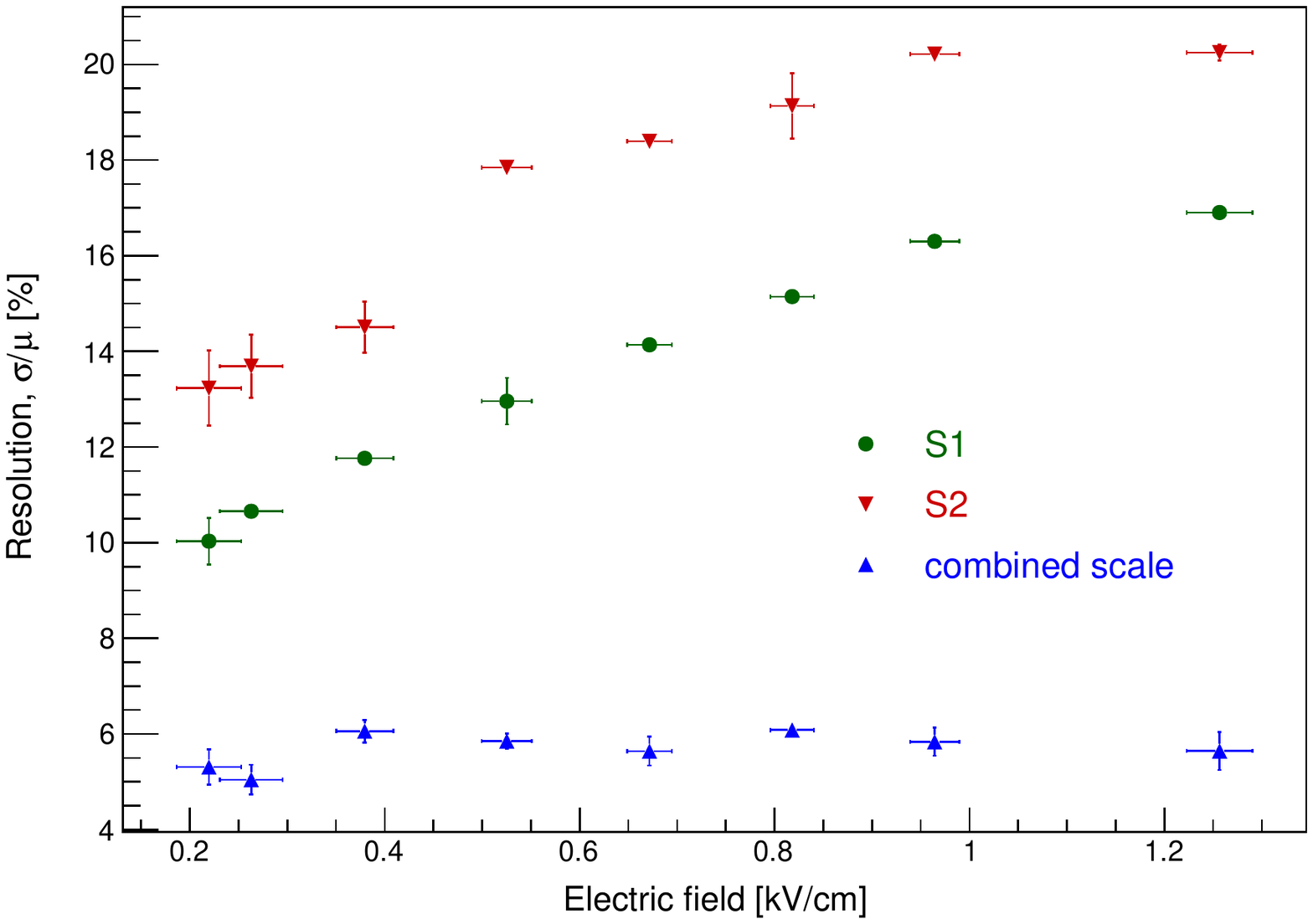}
\caption{\small Energy resolution for the 32.1\,keV line as a function of the electric drift field. While the resolution using only one of the two signal channels, S1 (green) or S2 (red)  deteriorates with increasing drift field, the resolution for the combined energy scale, $E_{\rm{CES}}$ (blue), remains constant as expected.}
\label{fig:calib_kr83m_s4}
\end{figure}

The energy resolution at 32.1\,keV is shown  as a function of the electron drift field in figure~\ref{fig:calib_kr83m_s4}. While the energy resolution using S1 or S2 signals alone deteriorates with increasing drift field, the resolution of the combined energy scale remains constant at (5.8$\pm$0.3)\%.

\subsection{Electron drift velocity measurements}
\label{sec:EDV}

The {\it{z}}-coordinate of an interaction in a two-phase TPC is linearly related to the measured  time delay between the prompt and proportional scintillation signals. This dependence, together with the knowledge of the physical dimensions of the TPC was exploited to measure the electron drift velocity as a function of the electric field in liquid xenon at a temperature of 184\,K. The thermal contraction of the PTFE (16.5$\times$10$^{-5}$~K$^{-1}$~\cite{PTFE_ThermalContraction}) has been taken into account, resulting in an absolute contraction of the TPC length by 0.46\,mm. The dominant uncertainty of 0.2\,mm is due to the tolerance in the machining of the structural components.

\begin{figure}[h!]
\centering
\includegraphics*[width=\columnwidth]{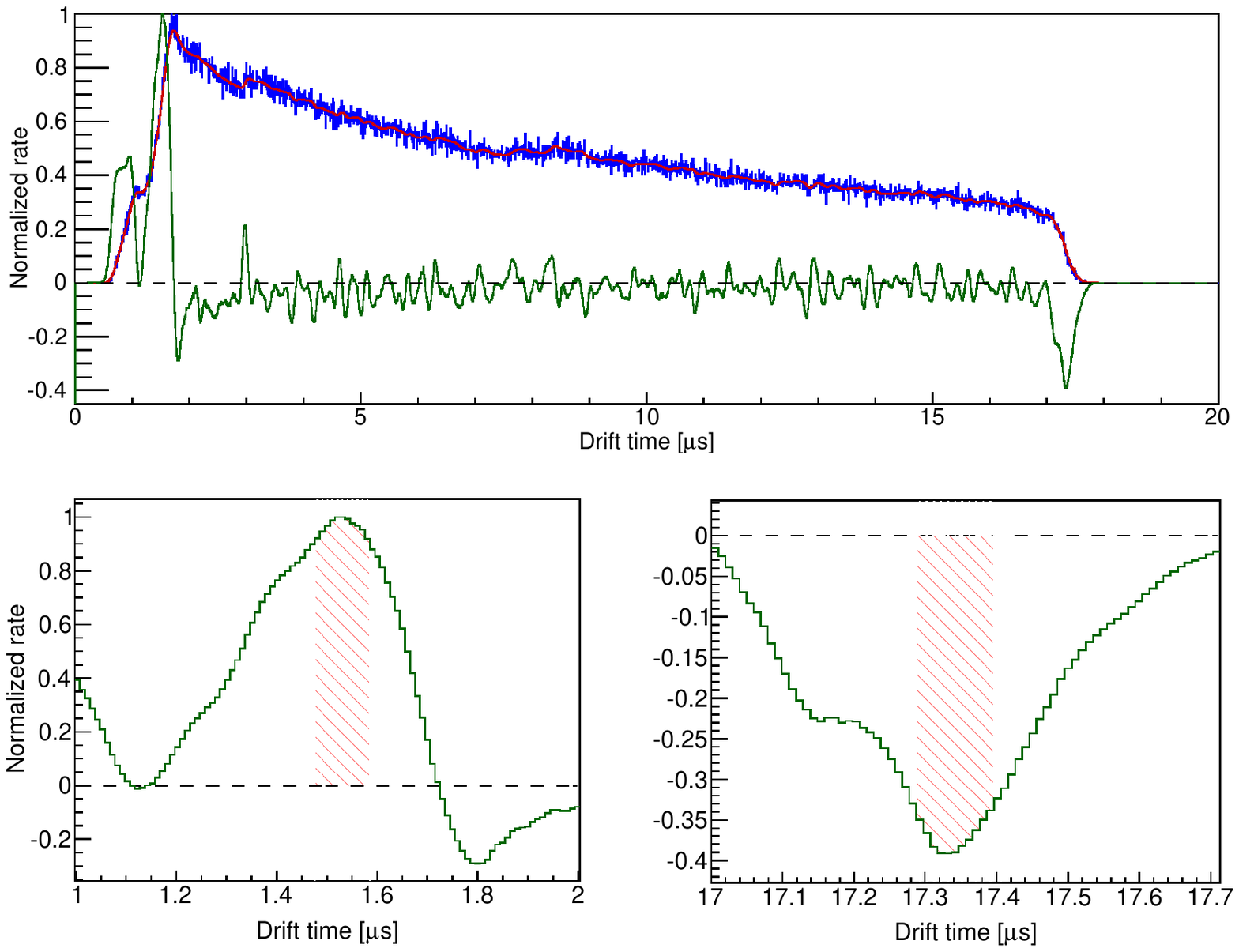}
\caption{\small ({\sl Top}): Drift time distribution for the 32.1\,keV line, normalised to the maximum value (blue histogram). Also shown is the convolution of the original distribution with a Gaussian kernel (red histogram), and the derivative of the smoothed drift time distribution, normalised to the absolute maximum (green histogram). ({\sl Bottom}): Zoom into the derivative of the time distribution, showing the time corresponding to the gate (left) and anode (right). The shaded region indicates the interval where the derivative is above 90\% of the absolute relative maximum value. This interval is taken as the uncertainty on the position of the two electrodes.}
\label{fig:DriftTimeDistr}
\end{figure}

The drift times corresponding to the positions of the gate and the cathode meshes were determined by the maximum and minimum of the time distribution of events, respectively, as demonstrated in figure~\ref{fig:DriftTimeDistr}.
To minimise the uncertainty on the drift time determination, we select events from  $^{83\text{m}}$Kr decays with the same time delay between the two largest S1 and largest S2 signals.
Since events related to background, dark counts in the PMTs or pileup are expected to have no correlation between the delay time of the two largest S1s with the delay time of the two largest S2s, this population shows a high purity of physical events related to $^{83\text{m}}$Kr decays.
Their distribution, shown in figure~\ref{fig:DriftTimeDistr}, was first smoothed with a Gaussian convolution kernel and then derivated. The drift times corresponding to the gate and cathode electrodes were therefore determined as the positions of the maximum and the minimum of the calculated derivative, respectively.

The systematic uncertainties were studied by employing a second method for the reconstruction of the electrode coordinates, based on the time delay of secondary electrons originating from the main S1 signal via photoionisation of the stainless steel meshes. The observed drift times, corresponding to the total drift length of the chamber, are in agreement within uncertainties with the results from the first method.

\begin{figure}[h!]
\centering
\includegraphics*[width=\columnwidth]{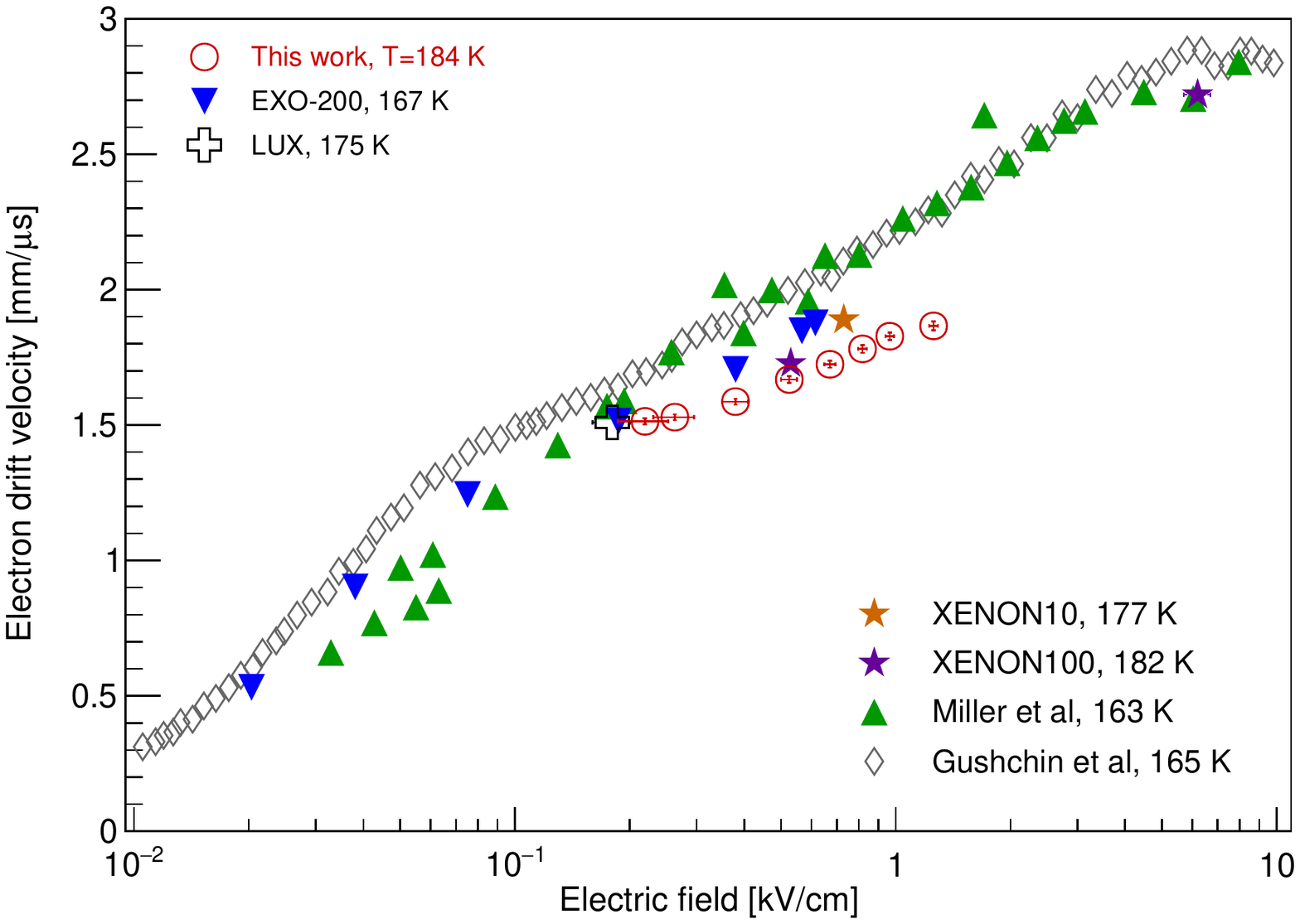}
\caption{\small Measured drift velocity of ionisation electrons as a function of electron drift field (open red circles) in the {\sl{Xurich\,II}} TPC, along with several literature values: XENON10 \cite{EDV_XENON10}, XENON100 \cite{Xe100_SingleElectrons,EDV_XENON100anal}, LUX \cite{EDV_LUX}, EXO-200 \cite{Albert:2016bhh}, as well as older measurements by Miller et al. at 163\,K \cite{EDV_Miller}, and Gushchin et al. at 165\,K \cite{EDV_Gushchin}.}
\label{fig:DriftVelocity}
\end{figure}

The measured electron drift velocity  as a function of electric field in the TPC is shown in figure~\ref{fig:DriftVelocity}, and compared to measurements by Miller et al. at 163\,K \cite{EDV_Miller}, Gushchin et al. at 165\,K \cite{EDV_Gushchin}, to measurements from XENON10 \cite{EDV_XENON10}, XENON100 \cite{Xe100_SingleElectrons,EDV_XENON100anal} and LUX \cite{EDV_LUX} at 177\,K, 182\,K, and 174\,K, respectively, and to the recent study by EXO-200 at 167\,K and fields below 0.6\,kV/cm \cite{Albert:2016bhh}. We find good agreement with the measurements by LUX and XENON, which were performed in a temperature regime similar to {\sl Xurich\,II}. The deviation from the EXO-200 data points, and in particular from older measurements by Miller and Gushchin, may be attributed to the known dependence of the electron mobility in LXe with the temperature, which roughly follows $\mu_{e}\propto T^{-3/2}$.

\section{Summary}
\label{sec:summary}

We described in detail a new small-scale, two-phase xenon TPC {\sl{(Xurich~II)}}, optimised for light and charge yield measurements at energies below 50\,keV. We characterised the performance of the TPC with calibration data acquired with an internal, uniformly distributed $^{83\mathrm{m}}$Kr source. The volume-averaged light yield at 9.4\,keV is 15.0\,PE/keV and 10.8\,PE/keV at zero and $\sim$1\,kV/cm electric drift field, respectively. It is 14.0\,PE/keV and 7.9\,PE/keV at an energy of 32.1\,keV. These light yields are higher than the ones obtained in the single-phase operation of the {\sl{Xurich~I}} detector~\cite{Manalaysay:2009yq}.
Together with an ionisation yield at these energies of 28 and 31 electrons/keV, respectively, and an S2 yield of 24\,PE/e$^-$,
we can reach a low energy threshold of 2-3\,keV in the nuclear-recoil equivalent energy scale, depending on the drift field.

The relative energy resolution, $\sigma/E$, using a linear combination of the light and charge signals is ($5.8\pm0.3$)\% at 32.1\,keV, comparable to other state-of-the-art, small-scale $(x,y)$-position sensitive liquid xenon detectors~\cite{MIX_TPC}.  The drift velocity of ionisation electrons was measured for electric fields from (0.22$\pm$0.03)\,kV/cm to (1.26$\pm$0.03)\,kV/cm and is in the range $(1.53-1.88)$\,mm/$\mu$s,  in agreement with literature values when we consider the temperature-dependence of the electron mobility.

Our near-future goals with this detector are to measure the electron extraction efficiency into the gas phase as a function of applied electric field, the nuclear versus electronic recoil discrimination as a function of drift field, as well as the light and charge yields of low-energy nuclear recoils generated by neutrons from a D-D fusion generator operated at the University of Z\"urich. We plan to improve the performance of the detector by replacing the top PMT with an array of segmented photosensors, thus adding $(x,y)$ position reconstruction capability.

\section*{Acknowledgements}
This work was supported by the Swiss National Science Foundation under Grants Nos. 200020-162501 and 200020-175863, by the European Union�s Horizon 2020 research and innovation programme under the Marie Sklodowska-Curie grant agreements No 690575 and No 674896, and  by the European Research Council (ERC) under the European Union's Horizon 2020 research and innovation programme, grant agreement agreement No 742789 ({\sl Xenoscope}). We  thank Andreas James for significant contributions to the design and construction of the TPC and its sub-systems, and Daniel Florin for his help in the design and production of the printed circuit boards and cabling of the detector.

\bibliographystyle{JHEP}
\bibliography{xurich_instrument_v1.9}

\end{document}